\documentclass[suppldata]{interact}

\usepackage[utf8]{inputenc}
\usepackage{amsmath,amsfonts,amssymb}
\usepackage{graphicx}
\usepackage{setspace}
\usepackage{booktabs}
\usepackage{eurosym}
\usepackage{siunitx}
\usepackage{adjustbox}
\usepackage[hyphens]{url}
\usepackage{hyperref}
\usepackage{nomencl}
\usepackage{natbib}

\makenomenclature

\newtheorem{remark}{Remark}

\title{A Comparative Analysis of Electricity Consumption Flexibility in Different Industrial Plant Configurations}
\author{
\name{Sebastián Rojas-Innocenti\textsuperscript{a}\thanks{Author Note: At the time of the study, Sebastián Rojas-Innocenti was affiliated with Fortia Energía. He is currently affiliated with Fundación CARTIF. CONTACT: Sebastián Rojas-Innocenti - sebroj@cartif.es.}, 
Enrique Baeyens\textsuperscript{b},
Alejandro~Martín-Crespo\textsuperscript{c},
Sergio~Saludes-Rodil\textsuperscript{c}
and Fernando~Frechoso\textsuperscript{d}}
\affil{
\textsuperscript{a}Fortia Energía, Gregorio Benítez 3--B, Planta 1, 28043 Madrid, Spain; 
\textsuperscript{b}Instituto de las Tecnologías Avanzadas de la Producción, Universidad de Valladolid, Paseo Prado de la Magdalena 3--5, 47011 Valladolid, Spain; 
\textsuperscript{c}CARTIF, Parque Tecnológico de Boecillo, Parcela~205, 47151 Boecillo, Spain; 
\textsuperscript{d}Departamento de Ingeniería Eléctrica, Universidad de Valladolid, Paseo Prado de la Magdalena 3--5, 47011 Valladolid, Spain}
}

\begin{document}

\maketitle

\begin{abstract}
The increasing integration of renewable energy sources into power systems is
intensifying the demand for greater flexibility among industrial electricity
consumers. However, operational constraints, production requirements, and
market dynamics pose significant challenges to achieving optimal flexibility.
This paper presents an enhanced mixed-integer linear programming (MILP) model
that directly optimizes electricity consumption flexibility in manufacturing
plants. Unlike previous approaches, the proposed model determines optimal
transactions with both day-ahead and intraday continuous electricity markets,
while ensuring production continuity and adhering to plant-specific operational
constraints. The methodology is validated through annual simulations of two
real-world industrial configurations—cement manufacturing and steel
production—using 2023 market data. Comparative results highlight that the steel
plant achieved average electricity cost savings through flexibility of
0.41~\euro/MWh, whereas the cement plant achieved 0.24~\euro/MWh, reflecting
differences in storage capacities, production rates, and operational
flexibility. A comprehensive sensitivity analysis further identifies key
parameters affecting flexibility potential, such as the production-to-demand
ratio, storage capacity, and minimum operation periods. The findings offer
valuable insights for industrial operators aiming to reduce energy costs,
enhance operational flexibility, and support the decarbonization of electricity
systems. 
\end{abstract}




\section{Introduction}
\label{sec:intro}

The accelerating integration of renewable energy sources into electricity grids demands urgent and effective flexible demand management strategies, particularly in industrial sectors characterized by high energy intensity and continuous production requirements. Industrial demand-side flexibility offers significant potential to enhance grid stability, reduce operational costs, and accelerate the decarbonization of energy systems.

However, optimizing industrial electricity consumption flexibility presents substantial challenges due to the complex interplay between operational constraints, production schedules, and the volatility of electricity markets. Although previous research has addressed various aspects of industrial demand response, few models simultaneously optimize participation across multiple electricity markets while rigorously considering real-world operational limitations.

To address these challenges, this paper proposes an enhanced mixed-integer linear programming (MILP) model specifically designed for industrial environments. The proposed methodology enables flexible production scheduling by optimizing electricity procurement strategies across both the day-ahead and continuous intraday markets, while strictly adhering to operational constraints such as minimum operating periods, storage capacities, and production continuity.

The model is validated through annual simulations of two real-world industrial configurations—a cement manufacturing plant and a steel production plant—using actual electricity market data from 2023. Additionally, a comprehensive sensitivity analysis identifies the key operational parameters that most significantly influence flexibility potential.

The primary contributions of this work are as follows:
\begin{enumerate}
\item Development of an enhanced MILP model that optimizes electricity
transactions in day-ahead and continuous intraday markets under realistic
operational constraints.

\item Validation of the model through annual simulations based on real-world
data from two distinct energy-intensive industries.

\item Comparative analysis of flexibility potential across different industrial
configurations.

\item Identification of key operational parameters affecting flexibility and
formulation of recommendations to optimize flexible industrial operations,
thereby offering practical insights for industrial operators and informing
energy policy development.  
\end{enumerate}

To guide the reader through the remainder of the manuscript, its structure is
organized as follows. Section~\ref{sec:Liter} reviews the existing literature
on industrial demand flexibility, with particular emphasis on the cement and
steel sectors. Section~\ref{sec:Metod} presents the methodological framework
and details the proposed optimization model. Subsequently,
Section~\ref{sec:Results} discusses the simulation results and the sensitivity
analyses conducted. Finally, Section~\ref{sec:Conc} summarizes the key findings
and proposes directions for future research.

\section{Literature Review}
\label{sec:Liter}

The integration of renewable energy sources into power systems and the
increasing complexity of electricity markets have intensified the need for
flexible electricity consumption strategies in industrial sectors. This section
reviews the existing literature on industrial demand flexibility, focusing on
the cement and steel industries due to their high energy intensity and
operational constraints. Additionally, other industrial sectors with
flexibility potential are discussed, culminating in the identification of key
research gaps addressed by this study.

\subsection{Industrial Demand Flexibility}

Industrial demand-side flexibility has been widely recognized as a crucial
mechanism for balancing supply and demand in increasingly volatile electricity
markets. Various studies have assessed the flexibility potential of industrial
processes and the role of demand response (DR) programs
\citep{rollert_demand_2022, pierri_integrated_2020}. Numerous optimization
approaches have been proposed to adjust industrial electricity consumption
according to price signals while preserving production schedules and
operational feasibility \citep{zhao_2014, boldrini_demand_2023}.

These contributions underscore the critical role that industrial sectors play
in enhancing grid stability. Building on this general perspective, the next
sections examine flexibility characteristics specific to the cement and steel
industries.

\subsection{Flexibility in the Cement Industry}

The cement industry exhibits significant flexibility opportunities,
particularly in processes such as raw milling and grinding, where operational
constraints are relatively less stringent compared to kiln operations. Several
studies have explored the participation of cement plants in DR programs by
shifting electricity consumption to off-peak periods
\citep{lee_evaluation_2020, olsen_opportunities_2011}. Optimization models
focused on minimizing electricity costs through load shifting and flexible task
rescheduling have been proposed \citep{olsen_opportunities_2011,
rombouts_flexible_2021, zhao_2014}, emphasizing the importance of production
flexibility in reducing operational expenses.

These findings highlight the importance of task rescheduling in cement
production for improving cost efficiency. To broaden the understanding of
sector-specific flexibility, the following section discusses flexibility
opportunities in the steel industry.

\subsection{Flexibility in the Steel Industry}

The steel industry, particularly facilities employing electric arc furnaces
(EAFs), has demonstrated considerable flexibility potential. EAFs allow
operators to adjust production schedules dynamically based on real-time
electricity market conditions, making them highly suitable for DR initiatives
\citep{boldrini_demand_2023, paulus_potential_2011}. Production scheduling
models have been developed to minimize electricity procurement costs under
price volatility, often integrating batch production strategies to enhance
operational flexibility \citep{marchiori_integrated_2017,
zhang_cost-effective_2017}.

This evidence suggests that technological and operational characteristics, such
as the use of EAFs, critically influence the ability of industries to
participate in electricity markets. Beyond these heavy industries, other
sectors have also shown notable flexibility potential.

\subsection{Flexibility in Other Industries and Residential Applications}

Other industrial sectors have exhibited substantial flexibility opportunities.
For example, the pulp and paper industry leverages batch production processes
to shift energy consumption \citep{arias2022demand}, while water treatment
facilities modulate pumping operations based on electricity price variations
\citep{torregrossa2016energy}. The food processing sector offers flexibility
through refrigeration systems, and plastics and rubber industries utilize
injection molding processes with adaptable load profiles
\citep{oldewurtel2015use, liu2022energy}. Research into the residential and
commercial sectors further provides insights into appliance-level flexibility
potentials \citep{rahmani_residential_2023, adiguzel_global_2024}.

While these studies affirm the widespread availability of flexibility across
diverse sectors, critical gaps persist, particularly regarding comprehensive
modeling approaches that capture operational realities.

\subsection{Research Gaps}

Despite the extensive research on demand flexibility and electricity cost
optimization across industrial sectors, significant limitations remain.
Specifically, few studies simultaneously optimize electricity procurement in
both the day-ahead and continuous intraday markets while rigorously considering
operational constraints such as storage limitations, minimum operation periods,
and production continuity. Additionally, many existing models idealize
flexibility potential without fully integrating the technical and operational
restrictions inherent to real-world applications.

This study aims to address these shortcomings by developing an enhanced
mixed-integer linear programming (MILP) model, explicitly designed to capture
the operational realities of cement and steel plant configurations, thereby
maximizing economic benefits through flexible electricity consumption.

\section{Methodology}
\label{sec:Metod}

This section presents the methodological framework employed to quantify
electricity consumption flexibility in industrial processes. It introduces the
problem context, describes the structure of the relevant electricity markets,
formulates the baseline and flexibility-enhanced optimization models, and
concludes with the case study configurations.

\subsection{Problem Statement}
\label{sec:problem_statement}

The primary objective of this study is to develop an enhanced procedure to
identify and quantify the electricity consumption flexibility of manufacturing
plants, building upon the methodology introduced in
\citep{rojasinnocenti2024electrical}. Several improvements have been
incorporated to refine and extend the original model:

\begin{itemize}

\item Direct determination of the optimal electricity quantities to purchase or
sell at each time interval, eliminating the need for exhaustive scenario
analyses.

\item Enabling participation in multiple transactions throughout the day within
the same market session, thus better exploiting intra-day price variability.

\item Enhancing computational efficiency to support faster simulations across
more complex scenarios.

\end{itemize}

Moreover, the model is designed to be applicable to a wide range of industrial sectors characterized by modular or batch-type production processes, where machinery can be operated flexibly in response to dynamic electricity prices while maintaining production continuity through intermediate material storage.

To assess the model’s performance, it is applied to two real-world industrial
configurations: a cement plant and a steel plant. 

While this study focuses on the cement and steel sectors due to the
availability of real operational data, the proposed methodology is applicable
to other industrial processes that share similar modular and batch-type
characteristics. For example, the pulp and paper
industry~\citep{arias2022demand}, water treatment
facilities~\citep{torregrossa2016energy}, and parts of the chemical
sector~\citep{romero2020flexibility} often employ flexible machines and
energy-intensive processes that can benefit from electricity consumption
optimization. Similarly, the food and beverage industry, particularly in cold
storage and dairy processing, has demonstrated potential for demand response
through flexible refrigeration and batching cycles~\citep{oldewurtel2015use}.
Plastic and rubber processing operations, such as injection molding, also
involve cyclic energy-intensive machinery that can be rescheduled to align with
electricity price signals~\citep{liu2022energy}. These similarities suggest
that the findings and modeling strategies developed here can offer relevant
insights for broader industrial applications beyond the case studies analyzed.

A comprehensive sensitivity analysis is also conducted to examine the impact of
key operational parameters on flexibility potential. Furthermore, the effects
of simultaneous variations in multiple parameters are analyzed to explore
possible synergistic or offsetting interactions.

\subsection{Electricity Market Structure}
\label{sec:market_chap}

This study considers the day-ahead market and the continuous intraday market
(SIDC) as the principal electricity procurement mechanisms. These markets
provide multiple trading opportunities throughout the day, allowing industrial
operators to adjust procurement strategies based on updated price signals and
operational conditions.

The Iberian electricity market, encompassing Spain and Portugal, is structured
into three segments: the day-ahead market, intraday auctions, and the
continuous intraday market (SIDC) \citep{OMIE_markets}.

The day-ahead market constitutes the primary channel for energy procurement, wherein participants submit bids covering the following 24-hour period. Market clearing is performed based on merit-order principles, producing an hourly price schedule.

By contrast, the SIDC facilitates continuous trading of electricity products across interconnected European regions up to one hour before delivery. Unlike intraday auctions, the SIDC enables real-time transactions and greater liquidity, offering flexibility to respond to forecast deviations and market developments.

Table~\ref{tab:SIDC_horario} summarizes the opening and closing times of the SIDC sessions relevant to the planning horizon of this study.

\begin{table}[htp]
\centering
    \begin{adjustbox}{width=0.70\linewidth}
        \begin{tabular}{cccll}
        \toprule
        \multicolumn{1}{c}{\bfseries Day} & \multicolumn{1}{c}{\bfseries\begin{tabular}{c}Contract\\starting time\end{tabular}} &\multicolumn{1}{c}{\bfseries\begin{tabular}{c}Contract\\end time\end{tabular}} &\multicolumn{1}{c}{\bfseries\begin{tabular}{c}Trading\\round\end{tabular}} &\multicolumn{1}{c}{\bfseries\begin{tabular}{c}SIDC\\negotiations periods\end{tabular}}\\
        \midrule
        D-1 & 14:00 & 15:00 & Round 17 & (D-1): 17..24\\
        \midrule
        D-1 & 15:00 & 15:20 & Round 18 & (D-1): 18..24\\
        D-1 & 15:20 & 16:00 & Round 18 & (D-1): 18..24  (D): 1..24\\
        \midrule
        D-1 & 16:00 & 17:00 & Round 19 & (D-1): 19..24  (D): 1..24\\
        D-1 & 17:00 & 18:00 & Round 20 & (D-1): 20..24  (D): 1..24\\
        D-1 & 18:00 & 19:00 & Round 21 & (D-1): 21..24  (D): 1..24\\
        D-1 & 19:00 & 20:00 & Round 22 & (D-1): 22..24  (D): 1..24\\
        D-1 & 20:00 & 21:00 & Round 23 & (D-1): 23..24  (D): 1..24\\
        D-1 & 21:00 & 22:00 & Round 24 & (D-1): 24  (D): 1..24\\
        \midrule
        D-1 & 22:20 & 23:00 & Round 1 & (D): 1..24\\
        \midrule
        D-1 & 23:00 & 0:00 & Round 2 & (D): 2..24\\
        D   &  0:00 & 1:00 & Round 3 & (D): 3..24\\
        D   &  1:00 & 2:00 & Round 4 & (D): 4..24\\
        D   &  2:00 & 3:00 & Round 5 & (D): 5..24\\
        D   &  3:00 & 4:00 & Round 6 & (D): 6..24\\
        D   &  4:00 & 5:00 & Round 7 & (D): 7..24\\
        D   &  5:00 & 6:00 & Round 8 & (D): 8..24\\
        D   &  6:00 & 7:00 & Round 9 & (D): 9..24\\
        D   &  7:00 & 8:00 & Round 10 & (D): 10..24\\
        D   &  8:00 & 9:00 & Round 11 & (D): 11..24\\
        D   &  9:00 & 10:00 & Round 12 & (D): 12..24\\
        D   &  10:00 & 11:00 & Round 13 & (D): 13..24\\
        D   &  11:00 & 12:00 & Round 14 & (D): 14..24\\
        D   &  12:00 & 13:00 & Round 15 & (D): 15..24\\
        D   &  13:00 & 14:00 & Round 16 & (D): 16..24\\
        \bottomrule
        \end{tabular}
    \end{adjustbox}
\caption{SIDC opening and closing times: Negotiating times depend on the specific time of day the market is accessed~\citep{OMIE_markets}.}
\label{tab:SIDC_horario}
\end{table}

In this framework, initial energy procurement is performed through the day-ahead market to establish a baseline schedule, while subsequent adjustments are considered via the SIDC. This two-stage procurement strategy enables the identification of additional flexibility opportunities while preserving operational feasibility.


\subsection{The Production Plant Model}

The flexible production sub-process is modeled as a combination of flexible production machines and product storage elements. This model is applicable not only to cement and steel production but also to various other industrial processes.

\begin{figure}[ht]
\centering
\includegraphics[width=.5\linewidth]{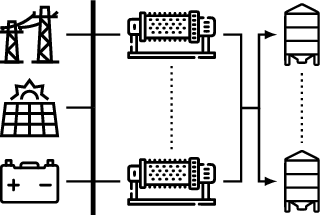}
\caption{The production plant model~\citep{rojasinnocenti2024electrical}.}
\label{fig:Plant}
\end{figure}

The machines, powered by electric motors, consume energy either from the grid or the plant’s self-consumption system, which includes photovoltaic panels and an electrical storage system. Figure~\ref{fig:Plant} provides a schematic representation of this production plant.

For a detailed explanation of the production plant model, the interested reader is referred to~\citep{rojasinnocenti2024electrical}. While the baseline model remains unchanged, several improvements have been introduced in the flexible model. Therefore, in the remainder of this section, we will describe only the constraints and cost functions required for the baseline model. A comprehensive explanation of the newly improved flexible model will be provided later in Section~\ref{sec:flex_prog}.

\paragraph*{Mass Balance.}

The mass balance in the production plant is given by
\begin{align}
\sum_{k\in\mathcal K} {\Pi_k}_t \cdot {Y_k}_t + 
\sum_{i\in\mathcal S} {I_i}_{t-1} = 
\sum_{i\in\mathcal S} {I_i}_{t}, 
\quad & t \in \mathcal T, \label{eq:mass_bal}
\end{align}

\paragraph*{Power Balance.}

The power balance in the production process is as follows,
\begin{align}
{P_b}_t + {P_D}_t + {P_\mathrm{PV}}_t = 
{P_s}_t + {P_C}_t + \sum_{k\in\mathcal K} Y_{kt} \cdot {P}_k, \quad
t \in \mathcal T.
\label{eq:power_bal}
\end{align}

\paragraph*{Silos Constraints.}

Let ${I_i}_{\min}$, ${I_i}_{\max}$, the minimum and maximum allowed limits of
silo $i\in\mathcal N$. The mass stored in each silo cannot exceed this limits:

\begin{align}
{I_{i}}_t \in [{I_{i}}_{\min}, {I_{i}}_{\max}].
\end{align}

The mass contained in the silo in each time slot  must be greater
than the product demand to ensure the continuity of the production process:

\begin{align}
\sum_{i\in\mathcal S} {I_i}_{t} \geq D_t
\end{align}
where $D_t$ is the product demand at time slot $t\in\mathcal T$.

\paragraph*{Machine Operation Constraints.}

Let $M_k^{\mathrm{ON}}$ be the number of time intervals that the machine k must remain on once it has changed its state from off to on.
Then, the inequality:
\begin{align}
\begin{aligned}
(Y_{k(t+1)} - Y_{kt}) \cdot M^{\mathrm{ON}}_k \leq \sum_{j=1}^{M^{\mathrm{ON}}_k} Y_{k(t+j)}, \\ 
k \in \mathcal K, \
t \in \lbrace1,\ldots,N_T-{M_k^{\mathrm{ON}}}\rbrace. 
\end{aligned}
\label{eq:PminON}
\end{align}
ensures that when the state of the machine changes from off to on, the machine remains on for $M_k^{\mathrm{ON}}$ time intervals.

Similarly, let $M_k^{\mathrm{OFF}}$ be the number of time intervals that the
machine $k\in\mathcal K$  must remain off once it has changed its state from
on to off.  
The inequality 
\begin{align}
\begin{aligned}
\sum_{j=1}^{M^{\mathrm{OFF}}_k} Y_{k(t+j)} \leq 
(1 + Y_{k(t+1)} - Y_{kt}) \cdot M^{\mathrm{OFF}}_k, \\
k \in \mathcal K, \
t \in \lbrace1,\ldots,N_T-{M_k^{\mathrm{OFF}}}\rbrace. 
\end{aligned}
\label{eq:PminOFF}
\end{align}
ensures that when the state of the machine changes from on to off, the machine remains on for $M_k^{\mathrm{OFF}}$ time intervals. 

\paragraph*{Battery Constraints.}

Considering that energy is the integral of power over time, and taking into account $\Delta t$ is the duration of the time slot during which power remains constant, we state the following inequalities
\begin{align}
\sum_{t=1}^{j} {P_C}_t \cdot \Delta t - \sum_{t=1}^{j} {P_D}_t \cdot \Delta t
\leq C_{\max} \cdot \mathrm{DoD} - \mathrm{SoC}_0, \quad
j \in \mathcal T \label{eq:max_SoC}\\
C_{\max} \cdot (1-\mathrm{DoD}) - \mathrm{SoC}_0
\leq
\sum_{t=1}^{j} {P_C}_t \cdot \Delta t - \sum_{t=1}^{j} {P_D}_t \cdot \Delta t,
\quad
j \in \mathcal T \label{eq:min_SoC}
\end{align}

Inequality~\eqref{eq:max_SoC} ensures that the battery charge never exceeds its rated capacity, while inequality~\eqref{eq:min_SoC} ensures that the battery is never fully discharged.

In addition, to preserve the health of the battery, the charge and discharge power cannot exceed a certain maximum value. This is ensured by the following conditions
\begin{align}
{P_C}_t \leq {P_C}_{\max}, \quad t \in \mathcal T \label{eq:max_Pcarga}\\
{P_D}_t \leq {P_D}_{\max}, \quad t \in \mathcal T \label{eq:max_Pdescarga}
\end{align}

Finally, a maximum value of electrical power ${P_b}_{\max}$ is allowed to buy
from the grid for each period in the given planning horizon.  
\begin{align}
{P_b}_t \leq {P_b}_{\max}, \quad
t \in \mathcal T.
\label{eq:limit_power_buy}
\end{align}

\subsection{The Optimal Production Schedule}

The production cost is defined as follows:
\begin{align} \label{eq:obj_func}
\text{Cost:} \ \Phi =
\sum_{t\in\mathcal T} \sum_{i\in\mathcal N} & 
\left(
{P_b}_t \cdot {\pi_b}_t +
( {P_C}_t + {P_D}_t ) \pi_U + I_{it} \cdot {\pi_S}_{it} 
\right) \cdot \Delta t.
\end{align}

\paragraph*{The Baseline Schedule.} It is the production plan that minimizes
production costs while meeting expected product demand over a given time 
horizon (typically one week in advance). It also satisfies all technical and product quality constraints. 
It is obtained by solving the following optimization program:
\begin{equation} \label{eq:baseline}
\begin{aligned}
\text{Minimize:}\   & \Phi,  \\
\text{subject to:}\ & \text{constraints}\ 
                       \eqref{eq:mass_bal}-\eqref{eq:limit_power_buy}, \\
\text{and:}\        & \text{non negativity for all variables.}
\end{aligned}
\end{equation}

The \emph{baseline schedule} is denoted as
\begin{align}
({P^*_b}_t,{P^*_C}_t,{P^*_D}_t,Y^*_{kt},
           {I^*_i}_t), \ i\in\mathcal S, k\in \mathcal K, t\in\mathcal T
\label{eq:blppx}
\end{align}
and the optimal cost is $\Phi^*$. 

\subsection{Flexibility in the Production Plan} 
\label{sec:flex_prog}

The ability of the manufacturing plant to provide flexibility to the electricity system is evaluated by perturbing the baseline schedule. Perturbing this schedule corresponds to the electricity that can be traded in the SIDC, achieved by selling energy previously purchased in the day-ahead market or by buying it when it was not initially acquired.

The modified production schedule is referred to as the \emph{flexibility schedule}, with production costs that are equal to or lower than those of the baseline schedule. The difference between production costs in the two scenarios is termed flexibility revenue. A positive flexibility revenue indicates profitable transactions in the intraday market, while a revenue of zero indicates that no profitable transactions are available.

The manufacturing plant operator has access to energy transactions only within a specific time horizon, determined by the opening and closing times of the SIDC. These times vary based on the time slot accessed, as discussed in Section~\ref{sec:market_chap}.

\subsubsection{The Flexible Schedule}\label{flexibility_model_chap}

Let $\mathcal T_1 = \{1,2,\ldots,N_{T_1}\}$ and $\mathcal T_2 = \{1,2,\ldots,N_{T_2}\}$ with \{$N_{T_1}, N_{T_2}\} < N_T$ be a subset containing the first $N_{T_1}$ and $N_{T_2}$ time slots of the production time horizon $\mathcal T$, respectively.
The time slots above mentioned represent the opening and closing times of the SIDC, which are determined by the specific time when the model is evaluated, denoted by $H_{SIDC}$. This symbol indicates the time slot when the model is queried. For more details, refer to Section~\ref{sec:market_chap}.

Let ${P_m}_t$ be the power purchased or sold from the SIDC at time interval $t\in\mathcal T$.  The variable is continuous and takes on negative values when selling what was previously purchased in the day-ahead market (${P_b}_t^*$) and positive values when purchasing.

The \emph{flexible schedule} is a perturbed production schedule of the baseline schedule where the perturbation is generated by a change in the power ${P_b}^*_{\tau}$ at time slot $\tau \in \mathcal T_1$ by the purchase or sale in the market SIDC represented by ${P_m}_t$. The flexible schedule is obtained by solving a new optimization program that has a similar cost function as the baseline schedule model, but with the additional term of power purchased multiplied by the price in the SIDC: 

\begin{align} \label{eq:obj_func2}
\begin{aligned}
\text{Cost:} \ \Phi^{\dag} = 
\sum_{t\in\mathcal T} \sum_{i\in\mathcal N} & 
\left(
{P_b}_t \cdot {\pi_b}_t + {P_m}_t \cdot {\pi_m}_t + \right. \\
& \left.  ( {P_C}_t + {P_D}_t ) \pi_U + I_{it} \cdot {\pi_S}_{it} 
\right) \cdot \Delta t
\end{aligned}
\end{align}

In addition, it should be noted that not all of the constraints change, since some of the decision variables keep the same value as in the baseline production plan. 

\paragraph*{Power Purchased from the Grid Constraints.}

In the flexibility model, the electric power purchased from the grid (${P_b}_t$) will take on different values than
the baseline schedule (${P_b}_t^*$) only after the time slot $\tau_2$, when the schedule can be rearranged by selling or buying energy only in the day-ahead market:

\begin{align}
{P_b}_t = {P_b}_t^*, & \quad  t > \tau_2, \ \tau_2 \in \mathcal T_2
\label{eq:same_grid}
\end{align}

\paragraph*{Power Purchased from the SIDC.} \label{MIC_purchase_var_chap}

This variable may only differ from zero between time slots designated as $\tau_1$ and $\tau_2$. These time slots are the only
ones allowed for the sale or purchase of energy due to the opening schedule of the SIDC. 

The resulting set of constraints are:
\begin{align}
\label{eq:same_MIC1}
{P_m}_t = 0, & \quad  t < \tau_1, \ t > \tau_2, \ \tau_1 \in \mathcal T_1, \ \tau_2 \in \mathcal T_2\\
|{P_m}_t| \leq LC1, & \quad \tau_1 \leq t \leq \tau_2, \ \tau_1 \in \mathcal T_1, \ \tau_2 \in \mathcal T_2
\label{eq:same_MIC2}
\end{align}

\paragraph*{Power Balance.}

The new power balance in the production process for the flexible schedule model is as follows:
\begin{align}
{P_b}_t + {P_m}_t + {P_D}_t + {P_\mathrm{PV}}_t = 
{P_s}_t + {P_C}_t + \sum_{k\in\mathcal K} Y_{kt} \cdot {P}_k, \quad
t \in \mathcal T.
\label{eq:power_bal2}
\end{align}

\paragraph*{The Flexible Schedule.}

It is obtained by solving the following Mixed Integer Linear Programming (MILP) optimization program:

\begin{equation} \label{eq:flexible}
\begin{aligned}
\text{Minimize:}\   & \Phi^{\dag} \text{defined in}~\eqref{eq:obj_func2}, \\
\text{subject to:}\ & \text{constraints}\ 
                      \eqref{eq:mass_bal}-\eqref{eq:limit_power_buy},
		      \eqref{eq:same_grid}-\eqref{eq:power_bal2}, \\
\text{and:}\        & \text{non negativity for all variables except for ${P_m}_t$.}
\end{aligned}
\end{equation}

The \emph{flexibility schedule} is denoted as
\begin{align}
({P^{\dag}_b}_t,{P^{\dag}_m}_t,{P^{\dag}_C}_t,{P^{\dag}_D}_t,
Y^{\dag}_{kt},{I^{\dag}_i}_t), \ 
i\in\mathcal S, \ k\in \mathcal K, \ t\in\mathcal T
\label{eq:fppx}
\end{align}
and the cost is $\Phi^{\dag}$.

The flexibility schedule is obtained by perturbing the baseline schedule, so its
cost is equal or lower than the cost of the baseline schedule, \emph{i.e.}
$\Phi^{\dag} \leq \Phi^*$. The cost of flexibility is defined as the difference
$\Delta\Phi^{\dag} =  \Phi^* - \Phi^{\dag}$
and is always positive or null in case of absence of recommended operations in the SIDC

\paragraph*{Price of Energy in the SIDC for Profitability.}
Flexible scheduling allows a certain amount of energy ${P_m}\cdot\Delta t \leq LC1\cdot\Delta t$ to be
available for trading in the SIDC. This energy can only be traded in this
market at the time interval between $\tau_1$ and $\tau_2$ which is the market
operation period. Trading is profitable
depending on the price of energy in the SIDC market and whether or not energy was purchased in the day-ahead market. Only two cases can occur:

\begin{enumerate}
\item[a)]
$\Phi^{\dag} < \Phi^*$, it means the existence of profitable transactions 
in the SIDC market, considering all readjustments in the day-ahead market after the time $\tau_2$. In this case, the quantity
$\Delta\Phi^{\dag}$ represents the revenue generated by perturbing the baseline schedule. This quantity must be greater than or equal
to the minimum revenue, denoted by R, that the plant operator must obtain to change the original optimal plan.

\item[b)] 
$\Phi^{\dag} = \Phi^*$, it indicates absence of profitable transactions in the SIDC market. In this case, $R = 0$, and there are no changes to the baseline schedule.
\end{enumerate}

\subsection{Industrial Case Study} \label{Case_study_main_chap}

This study presents a comparative analysis of the cement and steel industries, based on real-world plant configurations provided by industry operators. To ensure the robustness of findings, simulations will be conducted for each plant, utilizing actual operational data.

The following sections will detail the distinct manufacturing processes employed by each industry, highlighting the specific sub-processes selected for simulation. The criteria informing the selection of these sub-processes will also be discussed, with attention to their relevance and impact within the overall production framework.

It is interesting to note that energy storage solutions, specifically batteries, and renewable energy sources, such as solar panels, have been excluded from this study, as they are not part of the existing infrastructure at these plants. These technologies were thoroughly analyzed in a prior study and are, therefore, outside the scope of the present work.

Nevertheless, the findings of the previous analysis
\citep{rojasinnocenti2024electrical} reinforce the relevance of incorporating
renewable generation and storage technologies in industrial transformation
scenarios. Specifically, photovoltaic systems consistently reduced electricity
procurement costs by minimizing dependence on the grid, while battery storage
enhanced operational flexibility by enabling energy purchases during periods of
low market prices. Although the combined deployment of both technologies
further improved overall economic performance, the benefits were not always
proportional to the additional installed capacity. Notably, the optimization of
production planning in conjunction with these technologies led to significantly
greater economic gains compared to flexibility actions alone, with photovoltaic
systems exhibiting shorter average payback periods. These insights confirm the
strategic value of integrating such technologies in future industrial
flexibility frameworks, despite their exclusion from the present study due to
infrastructure limitations.

\subsubsection{Process Description}

\begin{figure}[h]
\centering
\includegraphics[width=.75\linewidth]{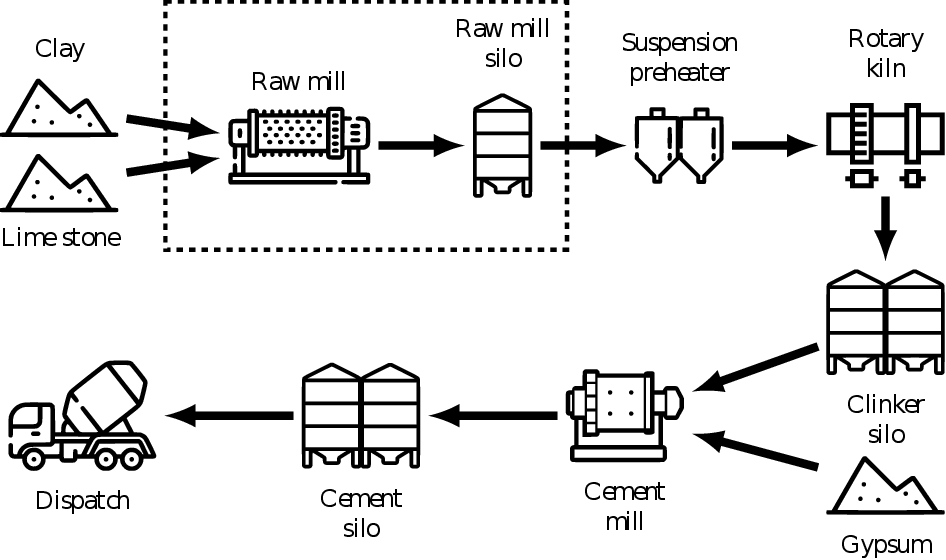}
\caption{Schematic representation of the Portland cement manufacturing~\citep{rojasinnocenti2024electrical}.}
\label{fig:CementProd}
\end{figure}

\paragraph*{Cement Manufacturing.}
An analysis of real data from the cement plant reveals that the raw mill production sub-process (highlighted by a dashed square in Figure~\ref{fig:CementProd}) exhibits the highest flexibility potential. This selection is primarily due to the sub-process’s comparatively lower production and quality constraints relative to other sub-processes. Consequently, it has been selected for simulation. For further details, refer to \citep{rojasinnocenti2024electrical}.

\paragraph*{Steel Production.}
Steel-making processes are classified into two main routes: the primary and secondary routes. The primary route produces steel from hot metal, using iron ore as the raw material in the initial reduction stage in a blast furnace~\citep{cavaliere_ironmaking_2016}. In contrast, the secondary route relies on scrap, sponge iron, or pig iron as inputs to produce steel~\citep{basic_concepts_2020_dutta}. According to the plant operator, the facility under analysis exclusively employs the secondary route, which will therefore constitute the sole focus of this description.

In the secondary steel-making route, scrap metal undergoes melting and decarburization in an Electric Arc Furnace (EAF). The crude steel is then transferred to a ladle, where primary alloying is typically conducted during tapping. The steel subsequently undergoes ladle treatment, which includes compositional adjustments, deoxidation, desulfurization, and degassing via vacuum treatment. Additional methods, such as gas rinsing or inductive stirring, are employed to enhance steel/slag interactions, remove deoxidation products, and achieve melt homogenization~\citep{holappa_chapter_2024}.

Following ladle treatments, the steel attains the specified composition and cleanliness, which must be preserved or potentially enhanced during the subsequent casting process. In contemporary continuous casting, steel is transferred from the ladle to a tundish and then into molds. This stage initiates the formation of a thin, solidified shell, setting the foundation for shaping the steel into various forms, such as flat sheets, beams, wires, or thin strips~\citep{holappa_chapter_2024}.

Primary forming continues this shaping process by employing hot rolling to refine the cast product, producing intermediate semi-finished forms---such as blooms, billets, and slabss---with precise dimensional and surface characteristics~\citep{basic_concepts_2020_dutta}.

The final phase, secondary forming, provides further shaping and property modifications through processes such as cold rolling, machining (e.g., drilling), joining (e.g., welding), coating, heat treatment, and surface finishing~\citep{basic_concepts_2020_dutta}. Within the plant under analysis, cold rolling is the sole secondary forming method utilized.

\begin{figure}[ht]
\centering
{\includegraphics[width=.75\linewidth]{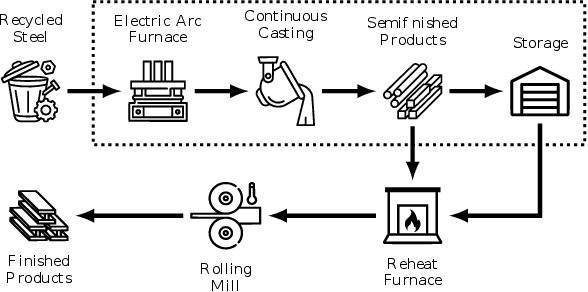}}
\caption{Schematic representation of the steel-making process~\protect\citep{sarda_multi-step_2021, steel_diag_website}}%
\label{fig:steel_diag}%
\end{figure}

Based on an evaluation of operational data from the plant, the study focuses on the melting phase sub-process, highlighted by the dashed rectangle in Figure~\ref{fig:steel_diag}, where an electric arc furnace (EAF) is employed.

The plant operator has indicated a feasible downtime window of at least one hour. However, to sustain continuous production, the melting process, along with continuous casting, must operate without interruption for a minimum of seven consecutive hours after the activation of the EAF.

Upon production of the semi-finished product, a portion is allocated to storage, while the remainder is directed to the next process, specifically the Reheat Furnace. This strategy ensures operational flexibility in the selected sub-process through the utilization of stored material. The operator has provided the average demand for the subsequent process, which will be used as a benchmark in our simulations. Further insights into the flexibility of this process are detailed in the following chapter.

\subsubsection{Demand Flexibility: A Comparison of Cement Manufacturing and Steel Production}

To evaluate the flexibility potential of the cement manufacturing and
steelmaking industries, two annual simulations were conducted using 2023
operational data. Each simulation incorporated the flexible machinery specific
to its respective sector, configured according to the operational constraints
defined by the plant operators. 

Although the analysis is limited to a single
year, the simulation spans the entire calendar of 2023, thereby capturing
seasonal fluctuations in electricity prices and production demand across
diverse operational conditions. Consequently, the results provide a
representative assessment of intra-annual dynamics and flexibility strategy
performance. Nonetheless, extending the study to encompass multi-year datasets
would improve the robustness of the findings by accounting for long-term market
trends, structural shifts, and atypical conditions, and thus constitutes an
important avenue for future research.

Each simulation began with the implementation of a baseline scheduling model
that utilized forecasted day-ahead market prices~\citep{sebastian2023adaptive}
to determine an optimal production schedule. This scheduling model operated
over a seven-day plus one (D-1) planning horizon, totaling 192 time slots per
week. Following this baseline, a flexible scheduling model was employed to
identify the optimal transactions based on actual SIDC market prices for that
year, as reported by OMIE~\citep{Precios_MIC_OMIE}.

Once both schedules were established, flexibility-induced savings were calculated as the cost differential between the two production schedules. This process was iterated daily over 365 cycles, simulating a full year for each industry. To ensure continuity, each day’s final material quantity was used as the initial quantity for the following day, thereby maintaining an ongoing production process throughout the year.

The following section provides a detailed discussion of the configuration parameters used in these simulations.

\begin{table}[htp]
\centering
    \begin{adjustbox}{width=0.80\linewidth}
        \begin{tabular}{cp{.4\linewidth}ccc}
        \toprule
        \multicolumn{1}{c}{\textbf{Parameter}} & \multicolumn{1}{c}{\textbf{Description}} &\multicolumn{1}{c} {\textbf{Cement}} &\multicolumn{1}{c} {\textbf{Steel}} &\multicolumn{1}{c} {\textbf{Units}}\\
        \midrule
        $P_t$                & Average electric power consumption of the flexible machine               & 6                                & 63                               & \unit{\MW\per\hour}\\
        ${\Pi}_t$            & Average production of the flexible machine                               & 360                              & 172                              & \unit{\tonne\per\hour}\\
        $D_t$                & Product demand for the next stage                                        & 240                              & 83.33                            & \unit{\tonne\per\hour}\\
        $M^{\mathrm{ON}}$    & Minimum hours of operation of the flexible machine                       & 6                                & 7                                & \unit{\hour}\\
        $M^{\mathrm{OFF}}$   & Minimum downtime of the flexible machine                                 & 3                                & 1                                & \unit{\hour}\\
        $I_{\max}$           & Maximum weight of material allowed in the storage                        & 15,000                           & 28,000                           & \unit{\tonne}\\
        $I_{\min}$           & Minimum weight of material allowed in the storage                        & $0.6 \cdot I_{\max} = 9,000$     & 0                                & \unit{\tonne}\\
        $I_{0}$              & Initial mass of material in the storage at the beginning of the week 0   & $0.6 \cdot I_{\max} = 9,000$     & 0                                & \unit{\tonne}\\
        $I_{n}$              & Initial mass of material in the storage at the beginning of the week n   & $I_{n} = I_{f}$                  & $I_{n} = I_{f}$                  & \unit{\tonne}\\  
        ${\pi_S}_{t}$        & Cost of storing material in the storage                                  & 0                                & 0                                & \unit{\EUR\per\tonne\hour}\\
        $H_{SIDC}$           & Time slot consulted at which the model is evaluated                      & 22                               & 22                               & \unit{\hour}\\
        $\tau_1$*            & Time slot for SDIC Opening                                               & 24                               & 24                               & \unit{\hour}\\
        $\tau_2$*            & Time slot for SDIC closing                                               & 48                               & 48                               & \unit{\hour}\\
        Battery**            & All the battery related parameters are null for this industrial cases    & -                                & -                                &  -            \\
        PV system**          & All the PV related parameters are null for this industrial cases         & -                                & -                                &  -            \\
        \bottomrule
        \end{tabular}
    \end{adjustbox}
\caption{Simulation Parameters for Analyzing Cement Manufacturing and Steel-making. *For more details, please refer to Chapter~\ref{sec:market_chap}. **Battery and PV system parameters are set to zero as these cases lack such installations.}
\label{table_parameters}
\end{table}

\paragraph*{Parameters Used for each Industries.} \label{param_simul_comparison}

The parameters applied in both scenarios are detailed in Table~\ref{table_parameters}, based on the actual configurations provided by each plant operator for the cement and steel manufacturing processes. These configurations reveal significant differences. For instance, the average electric power consumption ($P_t$) in the cement plant is ten times lower, while the production rate (${\Pi}_t$) in the cement plant is twice that of the steel plant. Conversely, the demand for the subsequent production stage ($D_t$) is twice as high in the cement plant compared to the steel plant. Although the minimum operating hours ($M^{\mathrm{ON}}$) are similar across both plants, the minimum downtime ($M^{\mathrm{OFF}}$) is considerably longer in the cement plant.

Regarding storage capacity, the steel plant’s maximum storage ($I_{\max}$) is twice that of the cement plant. Given that the minimum allowed storage ($I_{\min}$) is set to zero, the effective storage capacity in the steel plant is therefore greater than in the cement plant. However, the demand of the subsequent production process imposes a dominant constraint on storage flexibility.

All simulations were conducted using a standardized consultation time slot ($H_{SIDC}$), set to the 22nd time slot of each day. Consequently, the SIDC opening hours consistently occurred between time slots $\tau_1 = 24$ and $\tau_2 = 48$. Further details can be found in Chapter~\ref{sec:market_chap}.

\begin{remark}
Before conducting the simulations, it is important to highlight that the constraints for the cement plant are comparatively stricter than those for the steel plant in these cases. As a result, higher optimized costs and lower savings due to flexibility are expected for the cement plant compared to the steel plant. The production-to-demand ratio (${\Pi}_t/D_t$) for the subsequent process is more advantageous in the steel plant than in the cement plant. This trend is similarly reflected in the effective storage capacity and the minimum downtime requirements of the flexible machinery.
\end{remark}

In the following chapter, a detailed analysis and discussion of the simulation comparison results will be presented. Additionally, the concepts of ratios and effective storage introduced previously will be examined in greater depth.

\section{Results and Discussion} \label{sec:Results}

The simulations discussed in Chapter~\ref{Case_study_main_chap} were executed using PYSCIPOPT in Python 3.11.5. PYSCIPOPT is a Python interface for the SCIP Optimization Suite~\citep{SCIP_solver_Python}, a high-performance, non-commercial solver designed for a variety of mathematical optimization problems, including Mixed Integer Programming (MIP)~\citep{SCIP_solver}.

The annual simulation for the cement plant required a total runtime of 10.5 minutes, whereas the steel plant simulation was completed in 8.87 minutes.

In Subsection~\ref{schedule_comparison}, a weekly example is provided where flexibility savings were significant, illustrating the model’s scheduling of both baseline and flexible operations. Subsequently, Subsection~\ref{costs_comparison} offers a comparative analysis of production costs and flexibility between the two plants studied. Lastly, Subsection~\ref{sens_analysis} presents the results of the sensitivity analysis.

\subsection{Production Example for a Specific Week} \label{schedule_comparison}

\begin{figure}[ht]
\centering
{\includegraphics[width=.49\linewidth]{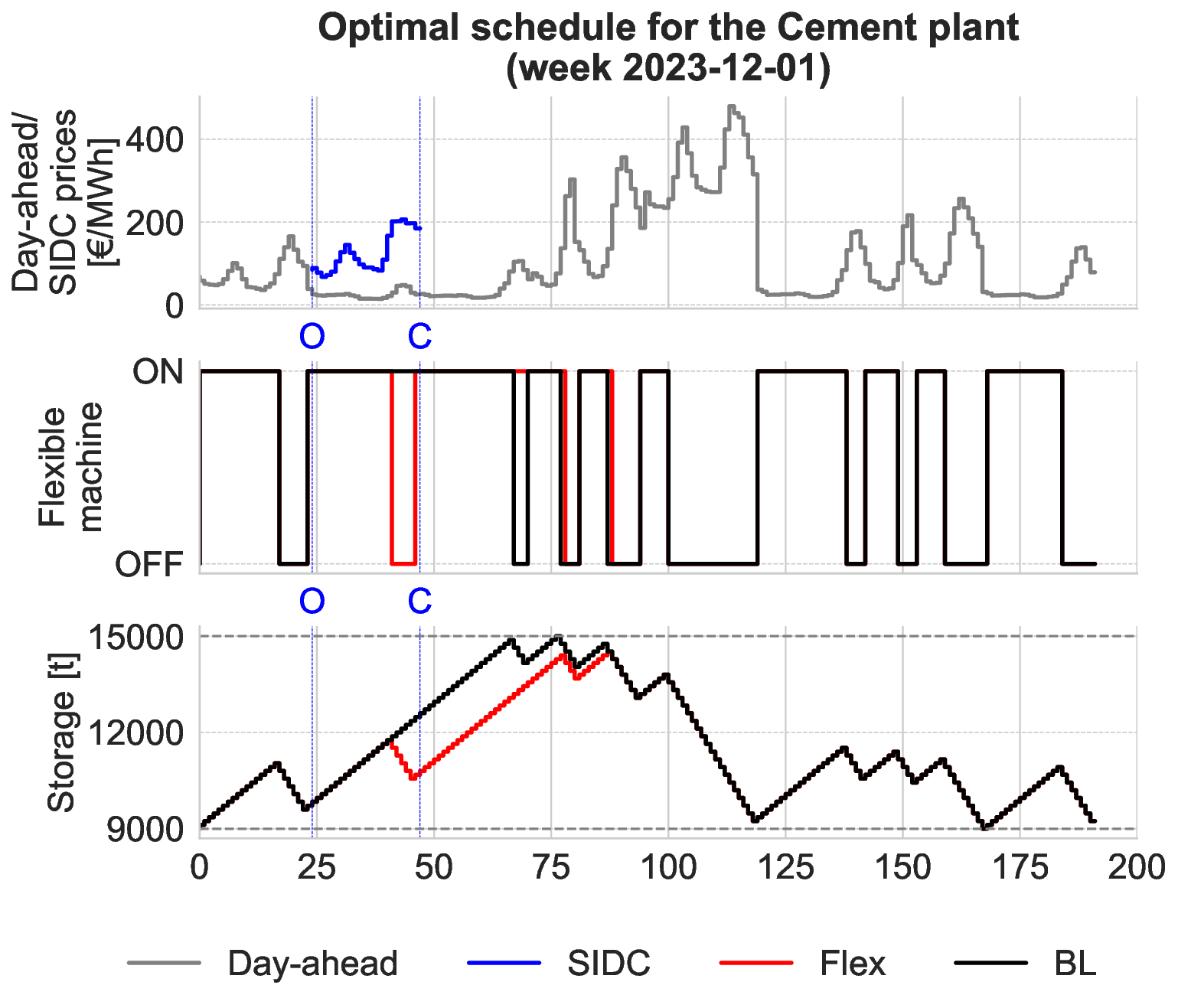}}
{\includegraphics[width=.49\linewidth]{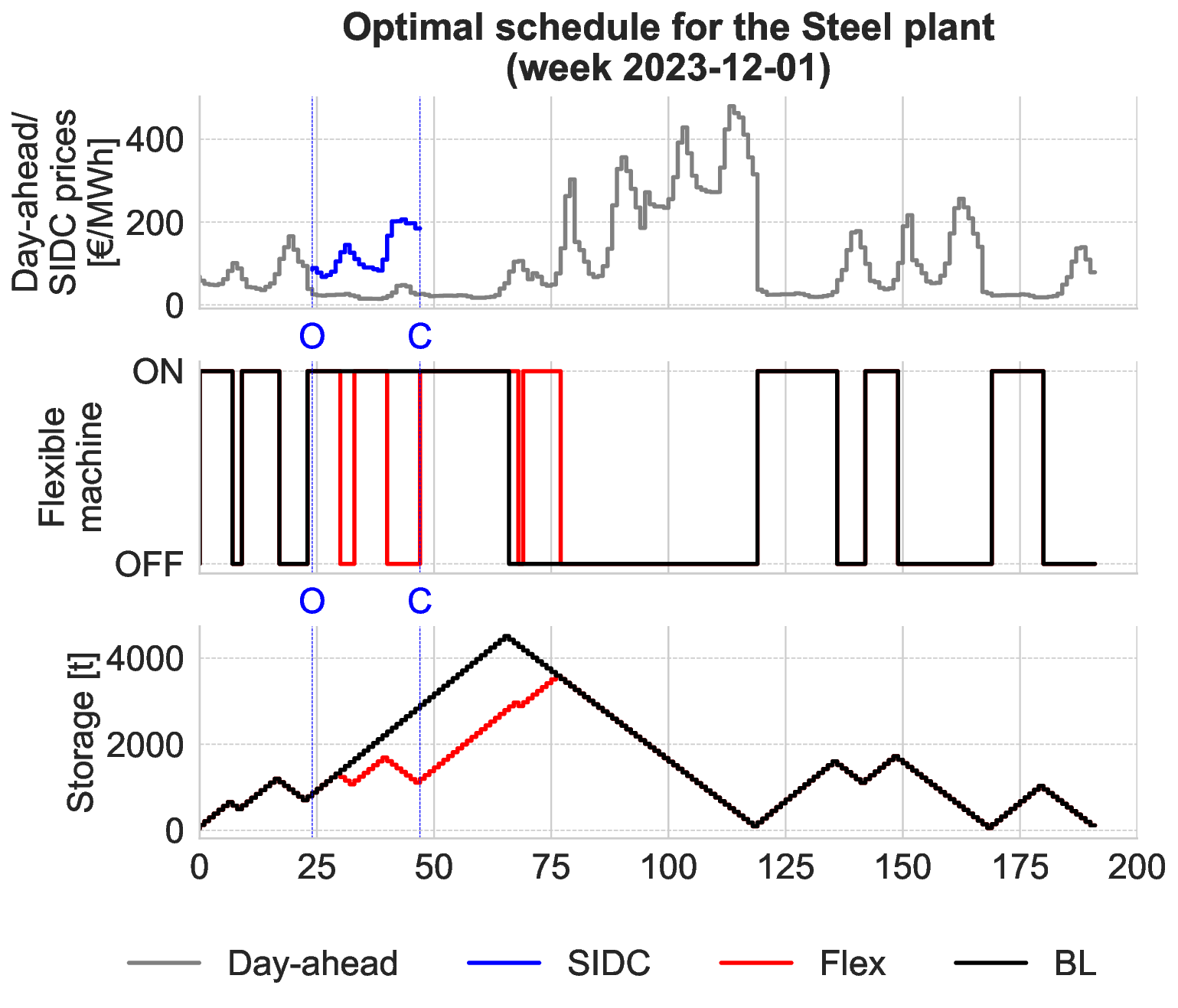}}
\caption{Comparison of optimal schedules for the Cement (left plot) and Steel (right plot) industries. The upper subplot shows day-ahead prices (grey) for the entire week alongside SIDC prices (blue) during market hours (with “O” and “C” indicating open and close, respectively, across 24–48 time slots). The middle subplot illustrates the optimal baseline schedule (black) and flexibility schedule (red) over a period of one week plus one day (192 time slots). The lower subplot displays material storage quantities, using the same color scheme.}
\label{fig:cem_vs_steel_week}%
\end{figure}

Figure~\ref{fig:cem_vs_steel_week} presents the simulation outcomes for both baseline and flexible scheduling in the cement plant (left plot) and the steel plant (right plot) for the week beginning on December 1, 2023. This week yielded particularly notable results in terms of flexibility.

The upper subplot illustrates the day-ahead prices in grey and the SIDC prices in blue, applicable exclusively during market opening hours (from the 24th to the 48th time slot, marked as “O” for opening and “C” for closing in each subplot). The middle subplot depicts the optimal baseline schedule (BL) in black and the flexibility schedule (Flex) as a red line, covering a full week plus one day (192 time slots). The lower subplot shows the amount of material stored throughout the week, employing the same color coding for each optimal scheduling scenario.

Although prices remain the same, the flexible scheduling of each machine differs considerably due to their distinct configurations and constraints.

The first remarkable feature is the behavior of the baseline schedule during the weekly price peak, which occurs approximately between time slots 75 and 120, as shown in the middle subplot. During this period, the steel plant successfully avoided energy purchases by utilizing stored material (indicated by the black line in the middle and lower subplots, respectively). In contrast, the cement plant was required to make purchases twice during these peak price intervals.

The second observation concerns the SIDC prices on Day D (represented by the blue line between the “O” and “C” markers in the upper subplot), which were significantly higher than the day-ahead prices (grey line). Consequently, the model seeks to maximize energy sales within the allowed constraints by selling energy initially purchased in the day-ahead market. The steel plant leveraged two SIDC price peaks, increasing its energy sales (depicted by the red line in the middle subplot), whereas the cement plant was limited to selling during the highest price peak.

Another remarkable observation is how the model readjusts the scheduling following the closure of the SIDC market (indicated by the blue line in the middle subplot after the “C” marker). These adjustments are made exclusively within the day-ahead market and are tailored differently for each plant to minimize electrical costs as effectively as possible.

It is also noteworthy that both plants began the initial day with stored material quantities close to the minimum permissible levels. Although these initial values were slightly above the absolute minimum due to the continuous nature of the simulation and the distinct configurations of each plant, they remained near the lower threshold. By the end of the week, storage levels in both plants were again close to minimum capacity. This outcome reflects the model’s cost-minimization approach, which operates the flexible machinery only as frequently as necessary to fulfill the hourly demand of the subsequent process.

On this exceptional day (December 1, 2023), the steel plant achieved flexibility savings of \qty{1056.59}{\EUR\per\MW}, while the cement plant reached only \qty{437.35}{\EUR\per\MW}. Similarly, the total annual flexibility savings amounted to \qty{11741.46}{\EUR\per\MW} for the steel plant and \qty{10742.35}{\EUR\per\MW} for the cement plant. This difference arises from the tighter constraints faced by the cement plant, which limited its transactions in the SIDC and restricted its readjustment opportunities within the day-ahead market.

Although these savings are expressed in \unit{\EUR\per\MW}, a direct comparison is not entirely accurate due to the differing constraints between industries. These constraints result in each plant operating their flexible machinery for varying durations over the year. Consequently, we have normalized the results to \unit{\EUR\per\MW\hour}. In the following section, a normalized comparison between the two plants is presented.

\begin{remark}
Note that the calculation of flexibility savings includes not only the revenue generated from energy transactions in the SIDC market but also the costs associated with adjusting the baseline schedule in the day-ahead market after the SIDC closure (following the 48th time slot or “C” marker). These adjustments are necessary to meet demand while maintaining the lowest possible electrical costs.
\end{remark}

\subsection{Production Costs and Flexibility Comparison} \label{costs_comparison}

To enable an accurate comparison between the two industries and their key parameter differences, production costs and flexibility savings were normalized to \unit{\EUR\per\MW\hour}. This normalization was performed by dividing the total annual costs by the power capacity of each flexible machine and the optimal total operational hours per year for each machine within each simulated scenario.

\begin{figure}[ht]
\centering
{\includegraphics[width=.5\linewidth]{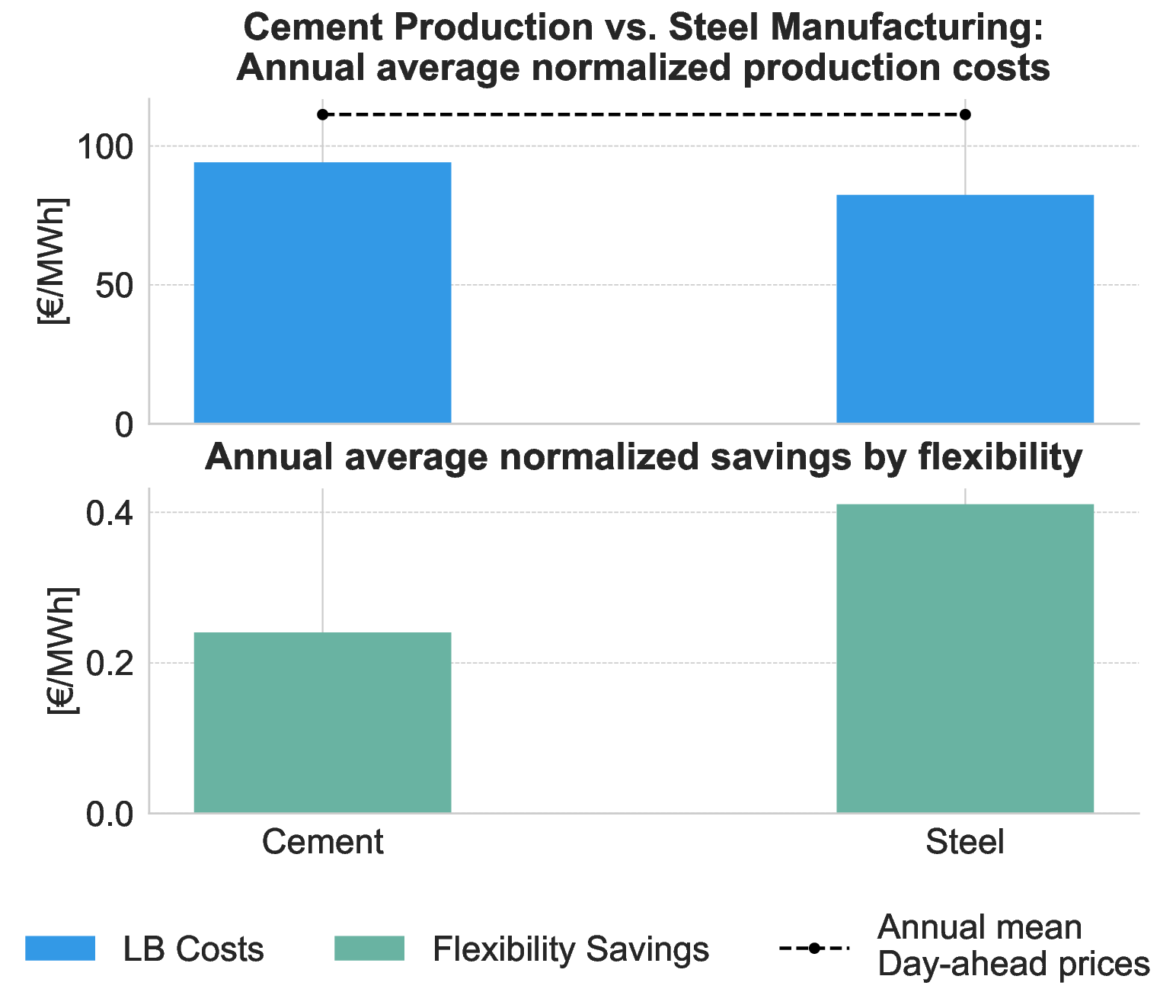}}
\caption{A comparative analysis of normalized production costs and flexibility savings between the cement and steel plants. In the upper plot, the blue bars indicate the normalized costs achieved through the optimal baseline scheduling strategy. The green bars in the lower plot, illustrate the difference between baseline and flexible normalized costs, representing the normalized flexibility savings.}
\label{fig:cem_vs_steel}%
\end{figure}

Figure~\ref{fig:cem_vs_steel} presents the simulation results. In the upper plot, a black dashed line represents the annual average day-ahead prices, which serves as a benchmark for assessing the model’s capacity to optimize production costs. The blue bars illustrate the normalized costs achieved by the optimal baseline scheduling strategy. In the lower plot, green bars depict the difference between baseline and flexible normalized costs, reflecting the savings realized through flexibility.

As expected, the steel plant demonstrates lower average normalized production costs compared to the cement plant. This result is due to several factors: a higher production-to-demand ratio for the flexible machine in the subsequent process, a larger usable storage capacity, and a longer minimum downtime requirement for the flexible machine. Together, these factors facilitate enhanced production optimization and flexibility savings in the steel plant.

These factors can impact production optimization and flexibility savings in different ways, with varying configurations yielding either positive or negative effects. Therefore, identifying these variables is crucial for understanding which ones have the most significant influence on optimization and flexibility.

To further investigate these effects, a sensitivity analysis was conducted, varying different parameters. The next section provides a detailed discussion of this analysis.


\subsection{Sensitivity Analysis} \label{sens_analysis}

A sensitivity analysis was performed on both the cement plant and the steel plant configurations. Parameters were adjusted to determine the optimal setup for each plant, with the primary objective of enhancing cost optimization and the secondary objective of increasing flexibility savings. The parameters that could be adjusted in the model were varied to observe their impact on electricity costs and flexibility savings, while always adhering to all previously described constraints, including total demand, maximum and minimum storage capacity, minimum operating hours, and minimum downtime of the machine.

\paragraph*{Demand as a Function of Flexible Machine Production.}

The most effective method for analyzing the demand of the subsequent process is to express it as a function of the production of the flexible machine, which can be represented by a production-demand ratio ($D_t/\Pi_t$). A higher ratio indicates that the demand closely resembles the production of the machine, implying a reduction in flexibility. This is due to the fact that the machine is unable to effectively manage the process output, which in turn makes scheduling downtime a more challenging task. Conversely, a lower ratio indicates that demand is considerably lower than production. This allows the machine to store excess production, thereby conferring the flexibility to power on and off freely during periods of low or high energy prices, respectively.

The maximum production-to-demand ratio applicable in the simulation without resulting in errors was 0.9 of the production capacity ($D_t = 0.9 \cdot {\Pi}_t = 0.9 \cdot \qty{360}{\tonne\per\hour} = \qty{324}{\tonne\per\hour}$). In contrast, the actual ratio observed in the cement plant is 0.67 ($D_t = 0.667 \cdot {\Pi}_t = 0.667 \cdot \qty{360}{\tonne\per\hour} = \qty{240}{\tonne\per\hour}$), while in the steel plant it is 0.48 ($D_t = 0.484 \cdot {\Pi}_t = 0.484 \cdot \qty{172}{\tonne\per\hour} = \qty{83.33}{\tonne\per\hour}$).

\begin{figure}[ht]
\centering
{\includegraphics[width=.55\linewidth]{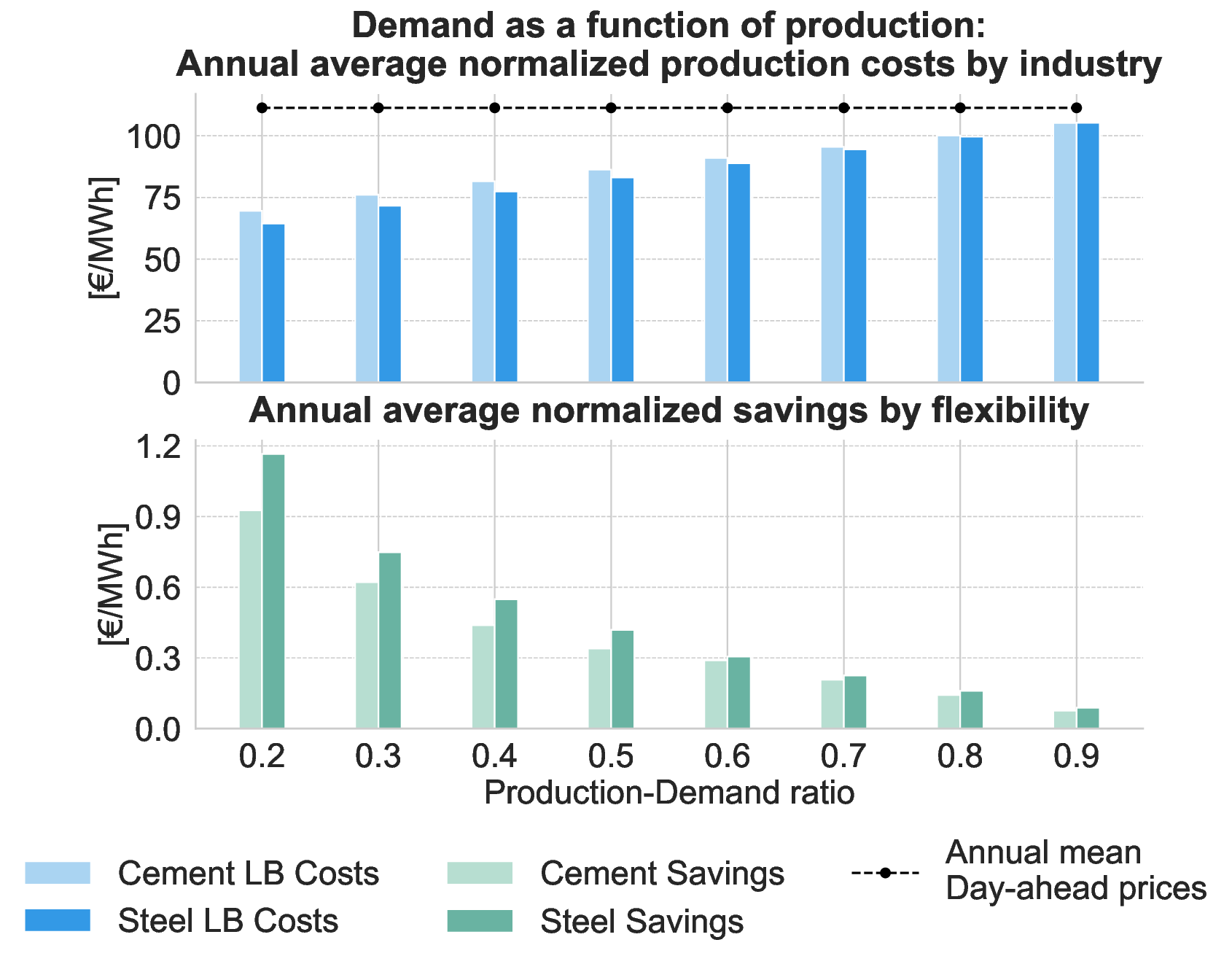}}
\caption{Iterating different demand values ($D_t$) as a function of flexible machine production (${\Pi}_t$), namely the production-demand ratio ($D_t/\Pi_t$). As the ratio increases, baseline normalized production costs rise, while the savings associated with flexibility decrease in both industries.}%
\label{fig:sens_demand}%
\end{figure}

Figure~\ref{fig:sens_demand} illustrates that normalized production costs increase as the production-to-demand ratio rises for both plant configurations. At a ratio of 0.9, where demand closely matches production, optimization potential is minimal in both cases. As a result, costs approach the annual day-ahead market average (black dashed line), indicating limited opportunities for the plant operator to leverage favorable energy prices while sustaining production and meeting demand. Similarly, flexibility savings decrease significantly as the ratio increases, reaching near-zero flexibility at a ratio of 0.9 for both configurations. Although the trend is consistent across both cases, the steel plant shows slightly higher flexibility savings. Normalized costs are generally lower for the steel plant across all evaluated values, except at a ratio of 0.9, where costs are marginally lower for the cement plant.

\paragraph*{Storage Capacity as a Function of Flexible Machine Production.}

As in our previous analysis, we evaluated the storage capacity relative to the production capacity of the flexible machine, represented by the storage-to-production ratio ($I_{\max}/{\Pi}_t$). This ratio reflects the number of times an hour’s worth of production can be stored in the facility. Higher ratios are expected to enhance flexibility, as the plant operator gains more opportunity to utilize stored material during periods of high electricity prices. Additionally, increased storage capacity allows for the accumulation of excess production when prices are low, enabling the machine to operate for extended periods under favorable pricing conditions.

However, as the ratio continues to rise, the potential for cost savings is limited, given that there is a maximum allowable amount of stored material that can be utilized within a one-week period.

\begin{remark}
A planning horizon of one week plus one day is consistently used, as price forecasts beyond this timeframe are not sufficiently reliable and would reduce the accuracy of the analysis.
\end{remark}

The minimum storage-to-production ratio that could be applied in the simulation without errors was 8 times the production capacity ($I_{\max} = 8 \cdot {\Pi}t = 8 \cdot \qty{360}{\tonne\per\hour} = \qty{2,880}{\tonne}$). The actual ratio observed in the cement plant is 41.67 times the production capacity ($I{\max} = 41.667 \cdot {\Pi}t = 41.667 \cdot \qty{360}{\tonne\per\hour} = \qty{15,000}{\tonne}$). It is important to note that usable storage does not equal maximum capacity; in this case, it is limited to $40\% \cdot I{\max} = 0.4 \cdot \qty{15,000}{\tonne} = \qty{6,000}{\tonne}$ due to a constraint that prevents storage from dropping below $I_{\min} = 60\% \cdot I_{\max} = \qty{9,000}{\tonne}$.

In contrast, the actual ratio observed in the steel plant is 162.79 times the production capacity ($I_{\max} = 162.791 \cdot {\Pi}t = 162.791 \cdot \qty{172}{\tonne\per\hour} = \qty{28,000}{\tonne}$). In this instance, the entire storage range is available for use, as there are no minimum storage constraints ($I_{\min} = 0$).

\begin{figure}[ht]
\centering
{\includegraphics[width=.55\linewidth]{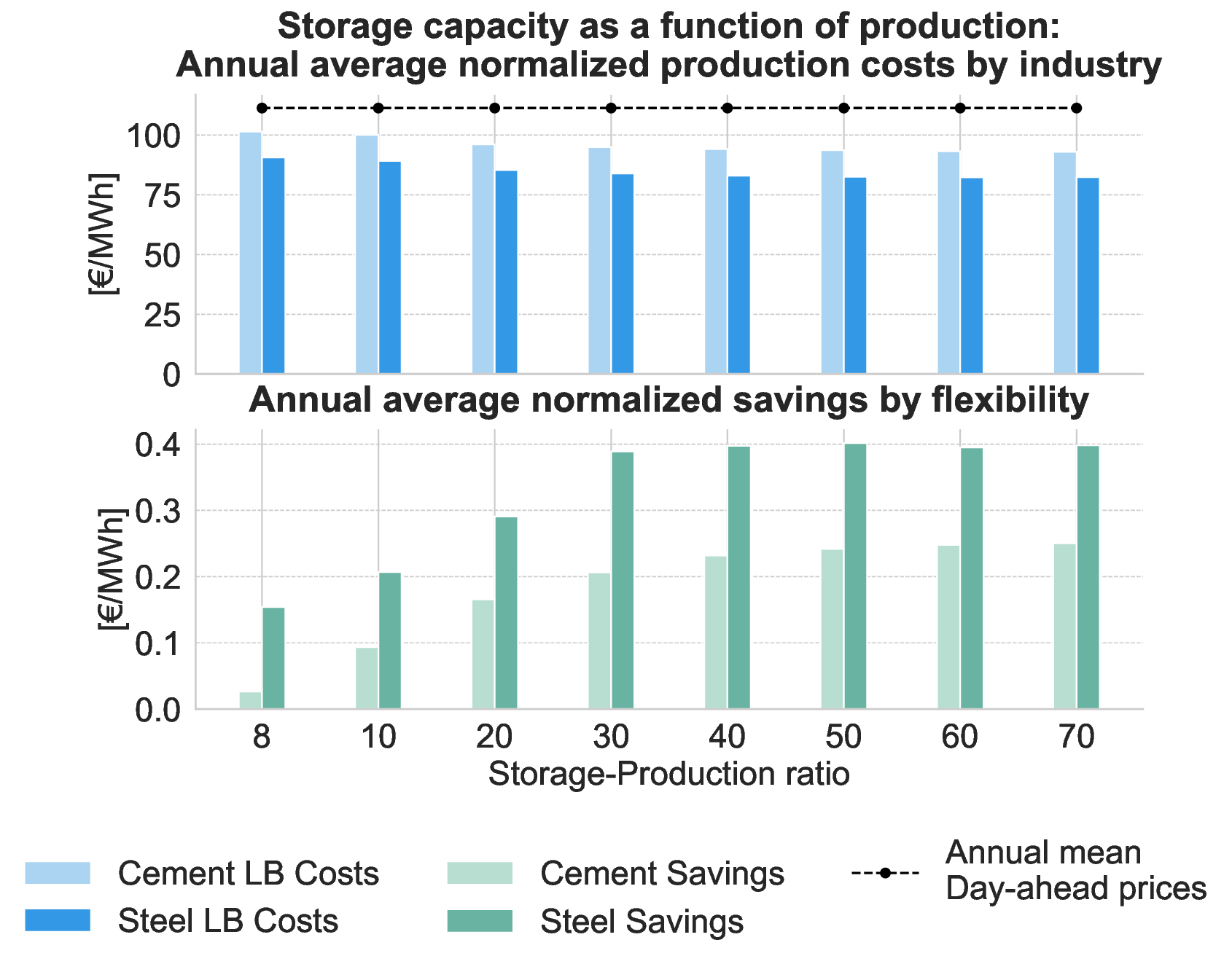}}
\caption{Iterating different storage capacity ($I_{\max}$) as a function of the production of the flexible machine (${\Pi}_t$), namely the storage-production ratio ($I_{\max}/{\Pi}_t$). In general, the normalized production costs decrease as the ratio increases, while the savings by flexibility increase. However, this is not a proportional relationship due to the surplus material stored being unable to be used effectively within a single week.}%
\label{fig:sens_silo}%
\end{figure}

Figure~\ref{fig:sens_silo} illustrates that a storage capacity equivalent to only eight times the output provides minimal flexibility in both plant configurations, with this effect being more pronounced in the cement plant. Interestingly, storage capacities of 50 to 70 times the output for the cement plant and 40 to 70 times for the steel plant yield constant flexibility savings in each case. This is because the large surplus of stored material cannot be fully utilized within a single week. In both cases, normalized production costs decrease as the storage-to-output ratio increases, though this decrease is not directly proportional due to the aforementioned limitations.

Consistent with previous results, both the normalized optimized costs and flexibility savings were more favorable for the steel plant than for the cement plant, due to the more advantageous constraints associated with these specific plant configurations.

\paragraph*{Minimum Operating Hours of the Flexible Machine.}

Another factor affecting flexibility is the minimum required operating time once the machine is switched on. A longer minimum operating time reduces flexibility, as it limits the model’s ability to activate the machine for short periods to benefit from lower energy prices. However, since lower prices often occur in consecutive intervals, the overall impact of this constraint may not be immediately significant.

\begin{figure}[ht]
\centering
{\includegraphics[width=.55\linewidth]{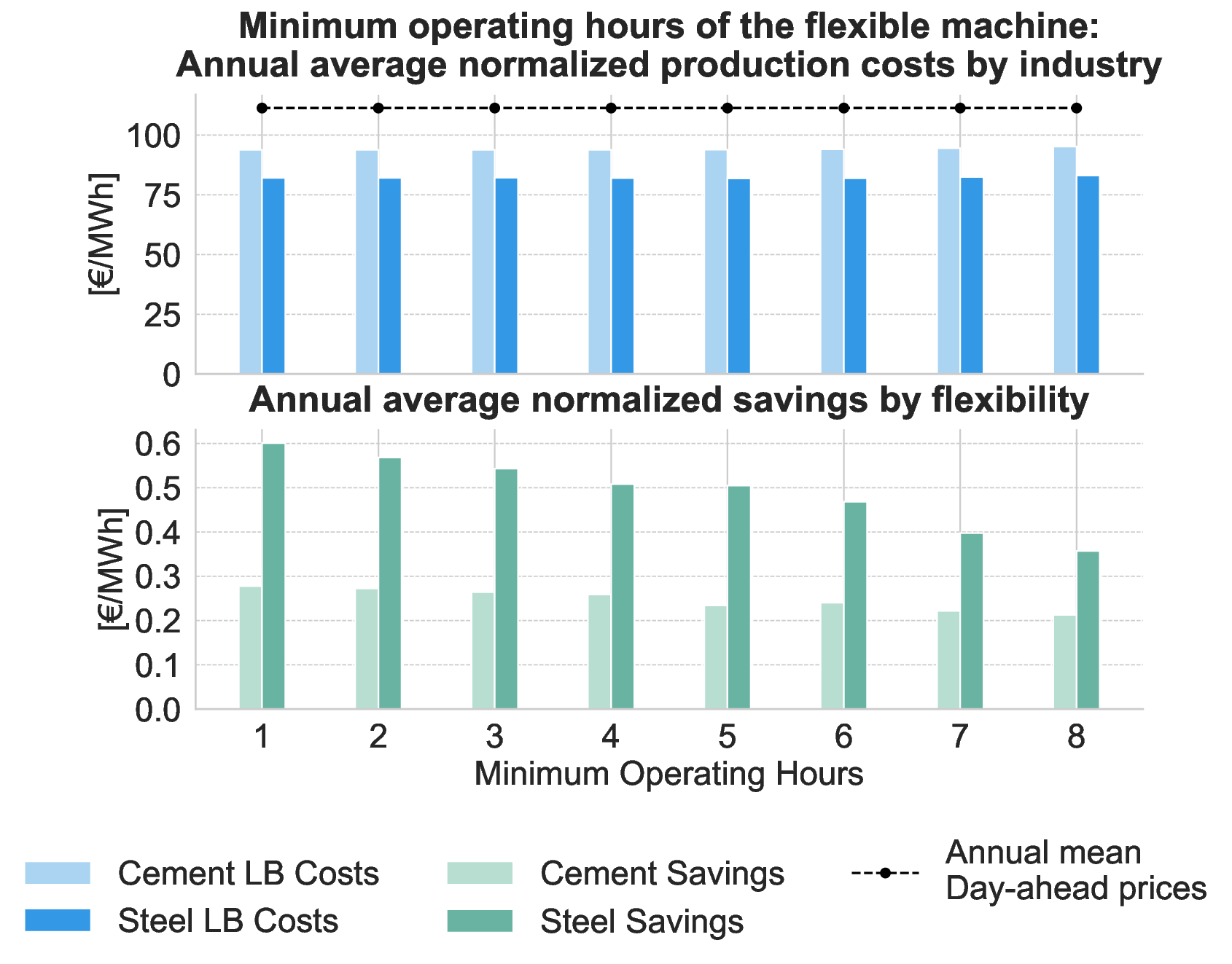}}
\caption{Iterating different Minimum Operating Hours ($M^{\mathrm{ON}}$) of the flexible machine. In general, shorter minimum operating time constraints result in greater normalized flexibility savings and lower normalized costs. However, the trend may not always be evident due to the tendency for lower prices to occur in succession.}%
\label{fig:sens_ON}%
\end{figure}

Figure~\ref{fig:sens_ON} illustrates that, overall, shorter minimum operating time constraints enhance flexibility savings in both plant configurations, though the trend is not always linear. This finding aligns with expectations, as consecutive periods of lower prices allow the machine to operate during these more cost-effective hours. Notably, in the cement plant, minimum operating constraints of 5, 7, and 8 hours yield the least favorable results, with a slight improvement observed at 6 hours. In contrast, constraints of 1 and 2 hours lead to optimal performance. For the steel plant, a clearer trend is observed, where stricter minimum operating time constraints result in reduced flexibility savings.

In the cement plant, normalized production costs consistently decline as the minimum operating time constraint relaxes; however, the magnitude of this decrease is very low. This pattern is not consistently evident in the steel plant, which aligns with expectations.

\begin{remark}
An additional simulation was conducted using a normally distributed random price dataset with the same mean and standard deviation as the original price dataset, serving as a control case. This control case eliminates the issue of consecutive low and high prices, allowing us to observe that with higher constraint hours for both minimum operating time and minimum downtime, optimization costs generally increased across all iterated values, while flexibility savings mostly decreased, though not in all instances. These results have been omitted due to space limitations.
\end{remark}

\paragraph*{Minimum Downtime of the Flexible Machine.}

As in the previous case, a shorter minimum downtime requirement after the machine is switched off leads to greater expected flexibility. This reduced constraint grants the model increased freedom to generate a more optimized schedule, allowing it to capitalize on fluctuations in energy prices. However, similar to the minimum operating time, the overall impact may not be entirely negative, as favorable and unfavorable prices often occur in succession.

\begin{figure}[ht]
\centering
{\includegraphics[width=.55\linewidth]{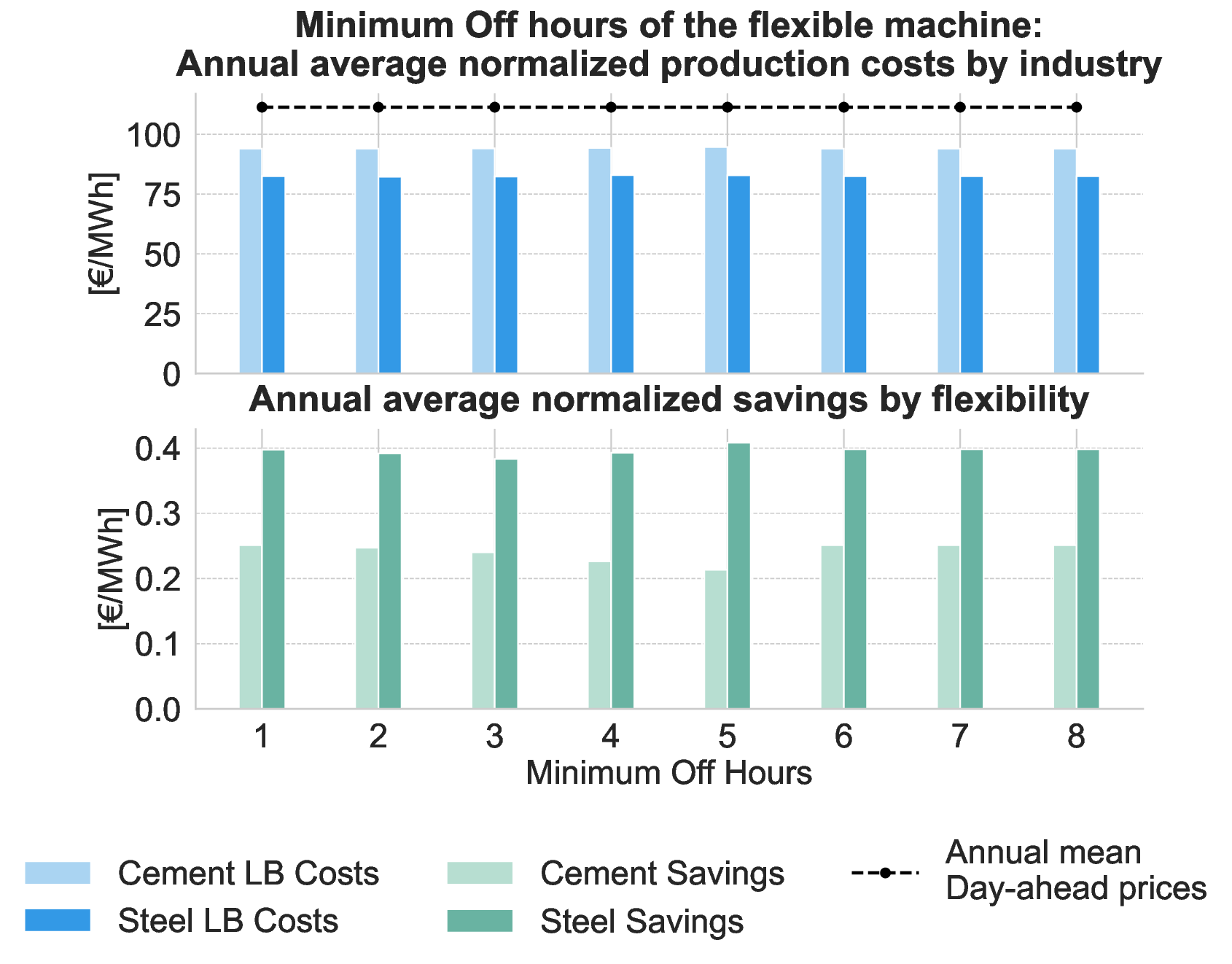}}
\caption{Iterating different Minimum Downtime ($M^{\mathrm{OFF}}$) of the flexible machine. The general trend indicates that the reduction in the number of required hours to maintain the machine off results in increased normalized flexibility savings and a decrease in normalized costs. However, these changes remain very low in magnitude.}%
\label{fig:sens_Off}%
\end{figure}

As illustrated in Figure~\ref{fig:sens_Off}, for the cement plant, the general trend suggests that shorter minimum downtime requirements lead to greater flexibility. However, this pattern is not consistent across all tested values. Constraints of 4 and 5 hours are the least advantageous, while 1, 2 and oddly 6, 7 and 8-hour constraints yield the best performance, showing similar normalized savings from flexibility. In the case of the steel plant, this trend is not observed, as the lowest flexibility appears with the 3-hour constraint. Regarding normalized production costs in the cement plant, an increase in cost is consistently observed as the constraints become more stringent, although these changes remain very low in magnitude. However, this behavior is less evident in the steel plant, where the lowest costs occur at 2 and 3-hour constraints. This further demonstrates that these constraints do not consistently improve results, for the reasons discussed above.

\paragraph*{Synergistic Evaluation of Optimal Parameters}

To assess whether the simultaneous application of the best-performing
parameters identified in the previous sensitivity analyses generates
synergistic effects on production cost and flexibility, a final simulation was
conducted incorporating all optimal values concurrently. The objective was to
determine whether their combined implementation would yield additional
improvements or reveal potential offsetting interactions.

The parameters selected for this evaluation were as follows:

\begin{itemize}

\item Production-to-Demand Ratio ($D_t/\Pi_t = 0.5$): An intermediate value was
chosen to balance oversized machine flexibility with realistic process demand,
improving upon the original ratios of 0.67 (cement) and 0.48 (steel).

\item Storage-to-Production Ratio ($I_{\max}/{\Pi}_t = 40$): A moderate
capacity was selected, lower than the original 41.67 (cement) and substantially
below 162.79 (steel), recognizing the diminishing returns of additional storage
beyond this threshold within a weekly planning horizon.

\item Minimum Operating Hours ($M^{\mathrm{ON}} = 1$): Reduced from the
original 6 hours (cement) and 7 hours (steel), this adjustment facilitates
shorter operational cycles.

\item Minimum Downtime ($M^{\mathrm{OFF}} = 1$): A unified value was selected
over the original 3 hours (cement) and 1 hour (steel), aiming to maximize
operational flexibility.

\end{itemize}

This configuration reflects a combination of improved and neutral adjustments.
For instance, the Minimum Downtime for the cement plant and the Minimum
Operating Hours for both plants were substantially optimized. Conversely, the
Minimum Downtime for the steel plant and the storage-to-production ratio for
the cement plant remained effectively unchanged, while slight compromises were
introduced in the steel plant’s storage-to-production ratio and the cement
plant’s production-to-demand ratio.

Consequently, this combined configuration allows the evaluation of whether the
cumulative benefits outweigh the minor trade-offs.

\begin{figure}[ht]
\centering
{\includegraphics[width=.6\linewidth]{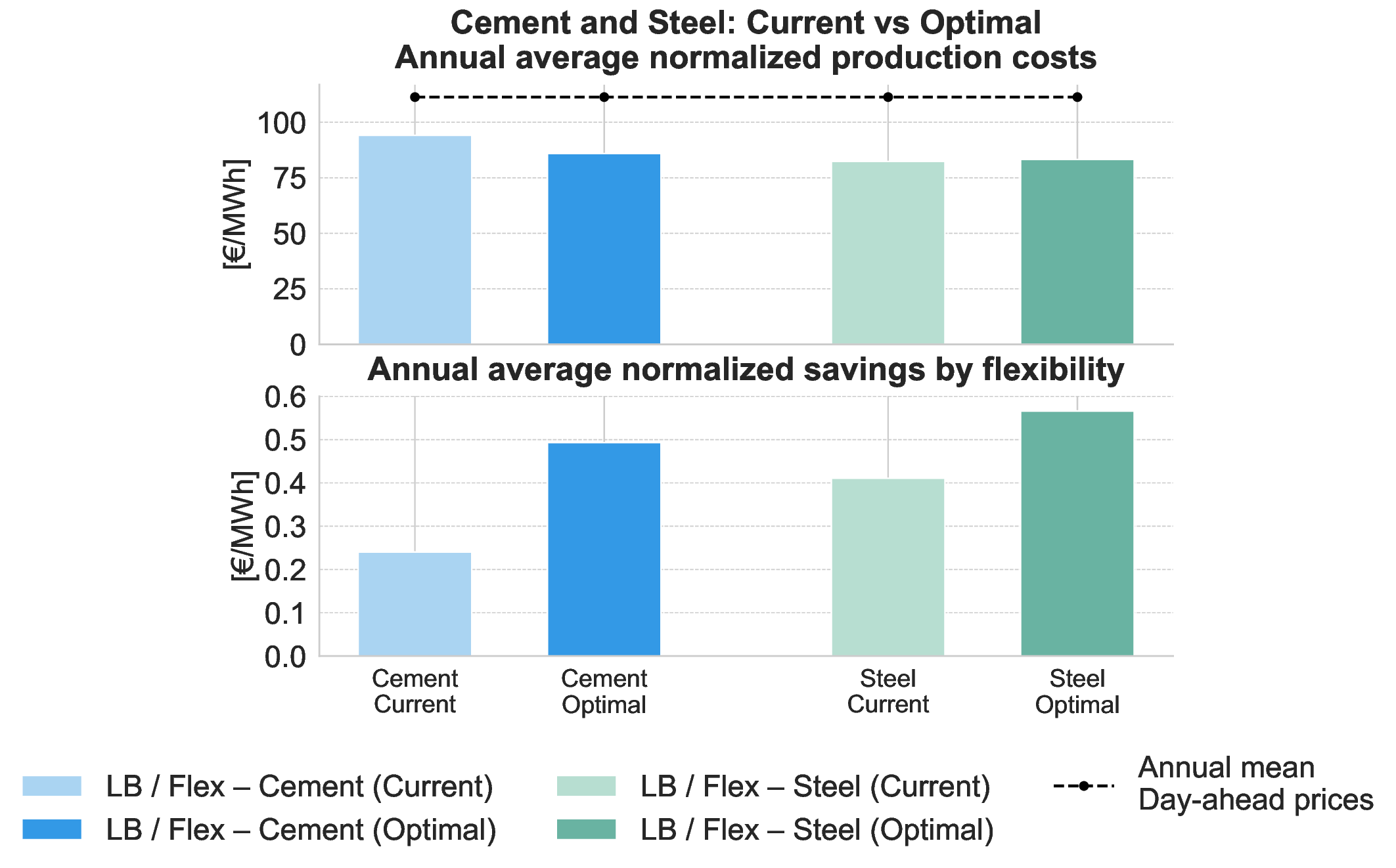}}
\caption{Comparison of normalized production costs and flexibility savings for the cement and steel plants, before and after applying the optimal set of parameters. The upper panel shows baseline normalized production costs, while the lower panel illustrates the normalized savings achieved through flexibility. The results highlight the cement plant's potential for further cost optimization and the overall improvement in flexibility for both configurations.}
\label{fig:cem_vs_steel_bef_after}%
\end{figure}

Figure~\ref{fig:cem_vs_steel_bef_after} presents the results of the synergistic
optimization. The upper panel illustrates the normalized baseline production
costs for the cement and steel plants, comparing the original and optimized
configurations. The cement plant exhibits a noticeable reduction in costs,
suggesting that its initial setup was suboptimal and benefited from parameter
adjustments. In contrast, the steel plant displays no significant change in
baseline costs, indicating that its original configuration was already
near-optimal in cost efficiency.

However, the most notable outcome is observed in the normalized flexibility
cost savings (lower panel), where both plants demonstrate substantial
improvements. This indicates that, even when baseline costs remain largely
unchanged (as in the steel plant), coordinated optimization of operational
parameters can significantly enhance electricity consumption flexibility.

\section{Conclusion}
\label{sec:Conc}

This study developed an enhanced mixed-integer linear programming (MILP) framework for optimizing industrial electricity consumption flexibility under realistic operational constraints. By explicitly modeling participation in both the day-ahead and continuous intraday electricity markets, the proposed approach enables the direct identification of economically optimal transaction strategies while preserving production continuity and technical feasibility.

The methodology was validated through annual simulations applied to two real-world industrial configurations: a cement manufacturing plant and a steel production facility. Comparative analyses revealed that the steel plant exhibited greater normalized flexibility savings relative to the cement plant, primarily attributable to differences in production-to-demand ratios, storage capacities, and minimum operating time requirements.

A comprehensive sensitivity analysis further identified the most influential
operational parameters—namely, the production-to-demand ratio, effective
storage capacity, and minimum operation periods—highlighting their critical
roles in shaping flexibility potential. Moreover, a synergistic evaluation
demonstrated that combined optimization across multiple parameters can yield
significant additional benefits, albeit with outcomes dependent on the specific
initial configuration of each plant.

Several limitations of this work must be acknowledged. First, the analysis was
confined to two industrial sectors, which may constrain the generalizability of
the findings. Nevertheless, the proposed methodology is readily adaptable to
other sectors characterized by modular or batch-type production processes, such
as the pulp and paper, water treatment, and chemical industries. Second,
simulations were conducted over a single calendar year, capturing seasonal
variations but not longer-term market dynamics; future research should thus
incorporate multi-year datasets to enhance robustness. Finally, the exclusion
of on-site renewable generation and storage systems, dictated by the current
infrastructure of the analyzed plants, suggests a valuable direction for future
investigations.

Future research will aim to extend this framework to multi-sectoral studies,
incorporate hybrid flexibility resources (including photovoltaic systems and
battery storage), and assess performance under high renewable energy
penetration scenarios. Furthermore, integrating flexibility optimization with
broader production planning and supply chain management considerations could
unlock additional synergies and economic efficiencies. Such extensions would be
particularly relevant for enhancing the resilience and economic viability of
industrial operations in future energy systems characterized by high levels of
variable renewable generation.

Overall, the findings of this study contribute to advancing the understanding
of industrial demand-side flexibility, offering practical insights for
industrial operators, electricity market designers, and policymakers seeking to
foster more resilient, cost-effective, and low-carbon energy systems.

\section*{Acknowledgments}
We appreciate the assistance of Fortia Energía for providing the related information on the Industrial Case Study.

\section*{Funding}
This research was founded by the MIG-20211033 grant from the Center for Industrial Technological Development (CDTI) of the Ministry of Science, Innovation, and Universities of the Spanish Government.
It has also been supported by Fundación CARTIF, a private and non-profit multidisciplinary Research Institution.

\section*{Conflict of interest statement}
The authors and Fundación CARTIF have no relevant financial or non-financial interests to disclose.

\section*{Data availability statement}
Data supporting the findings of this study are available from the corresponding author, SRI, upon reasonable request.

\section*{Author contributions statement}
\textbf{Sebastián Rojas-Innocenti:} Conceptualization, Data Curation, Formal analysis, Investigation, Methodology, Resources, Software, Visualization, Validation, Writing - Original Draft, Writing - Review \& Editing.
\textbf{Enrique Baeyens:} Conceptualization, Formal analysis, Investigation, Methodology, Resources, Supervision, Validation, Visualization, Writing - Review \& Editing.
\textbf{Alejandro Martín-Crespo:} Conceptualization, Funding acquisition, Methodology, Project administration, Resources, Supervision, Validation, Writing - Review \& Editing, Visualization.
\textbf{Sergio Saludes-Rodil:} Conceptualization, Funding acquisition, Methodology, Project administration, Validation, Resources, Writing - Review \& Editing, Supervision. 
\textbf{Fernando Frechoso-Escudero:} Conceptualization, Methodology, Resources, Supervision, Writing - Review \& Editing, Visualization.

All authors have read and approved the final version of the manuscript.

\nomenclature[V]{${P_b}_t$}{Power purchased from the grid in the period $t$ [{\unit[per-mode = symbol]{\MW}}].}
\nomenclature[V]{${\pi_b}_t$}{Day-ahead energy price forecast for period $t$ [{\unit[per-mode = symbol]{\EUR\per\MW\hour}}].}
\nomenclature[V]{${P_C}_t$}{Power used to charge the battery in the period $t$ [{\unit[per-mode = symbol]{\MW}}].}
\nomenclature[V]{${P_D}_t$}{Power obtained from discharging the battery in the period $t$ [{\unit[per-mode = symbol]{\MW}}].}
\nomenclature[V]{${\pi_S}_{it}$}{Cost of storing material in the $i$-th silo from one period to the next [{\unit[per-mode = symbol]{\EUR\per\tonne\hour}}].}
\nomenclature[V]{$Y_{kt}$}{Binary variable that represents the ON/OFF state of the $k$-th machine in the time t.}
\nomenclature[V]{${\Pi_k}_t$}{Average production of the $k$-th machine in the period $t$ [{\unit[per-mode = symbol]{\tonne\per\hour}}].}
\nomenclature[V]{${I_i}_t$}{Mass weight of the material stored in the $i$-th silo in the period $t$ [{\unit[per-mode = symbol]{\tonne}}].}
\nomenclature[V]{$D_t$}{Average product mass flow demand needed for the next process at the period $t$ [{\unit[per-mode = symbol]{\tonne\per\hour}}].}
\nomenclature[V]{${P_m}_t$}{Power purchased or sold from the SIDC at time interval $t$ [{\unit[per-mode = symbol]{\MW}}].}
\nomenclature[V]{${\pi_m}_t$}{The price signal for buying or selling electrical energy in the SIDC market during the time period $t$. [{\unit[per-mode = symbol]{\EUR\per\MW\hour}}].}
\nomenclature[C]{$\tau_1$}{The opening time slot for the SDIC is dependent on the value of $H_{SDIC}$ [{\unit[per-mode = symbol]{\hour}}].}
\nomenclature[C]{$\tau_2$}{The closing time slot for the SDIC is dependent on the value of $H_{SDIC}$ [{\unit[per-mode = symbol]{\hour}}].}

\nomenclature[C]{$\pi_U$}{The battery cost per unit of energy. Is the same for charging and discharging, and is a constant value defined by the battery's technical characteristics [{\unit[per-mode = symbol]{\EUR\per\MW\hour}}].}
\nomenclature[C]{$P_k$}{Average power consumption of the $k$-th machine [{\unit[per-mode = symbol]{\MW}}].}
\nomenclature[C]{${P_b}_{\max}$}{Maximum power purchase limit [{\unit[per-mode = symbol]{\MW}}].}
\nomenclature[C]{$C_{\max}$}{Battery rated capacity [{\unit[per-mode = symbol]{\MW\hour}}].}
\nomenclature[C]{$\mathrm{DoD}$}{Battery depth of discharge, is the fraction of the battery's rated capacity that can be discharged. It is a parameter given by the battery manufacturer [{\numrange{0}{1}}].}
\nomenclature[C]{$\mathrm{SoC}_0$}{Initial battery state of charge at the beginning of a given time planning horizon [{\unit[per-mode = symbol]{\MW\hour}}].}
\nomenclature[C]{${P_C}_{\max}$}{Battery maximum charge power [{\unit[per-mode = symbol]{\MW}}].}
\nomenclature[C]{${P_D}_{\max}$}{Battery maximum discharge power [{\unit[per-mode = symbol]{\MW}}].}
\nomenclature[C]{${I_{\max}}_i$}{Maximum weight of material allowed in the $i$-th silo [{\unit[per-mode = symbol]{\tonne}}].}
\nomenclature[C]{${I_{\min}}_i$}{Minimum weight of material allowed in the $i$-th silo [{\unit[per-mode = symbol]{\tonne}}].}
\nomenclature[C]{${I_{0}}_i$}{Initial weight of material in the $i$-th silo at the start of the planing horizon [{\unit[per-mode = symbol]{\tonne}}].}
\nomenclature[C]{$M_k^{\mathrm{ON}}$}{Minimum number of periods the $k$-th machine must operate once turned on for technical or quality reasons given by the plant.}
\nomenclature[C]{$M_k^{\mathrm{OFF}}$}{Minimum number of periods the $k$-th machine must remain turned off once is switched off for technical or quality reasons given by the plant.}
\nomenclature[C]{$H_{SIDC}$}{Specific time slot at which the model is evaluated [{\unit[per-mode = symbol]{\hour}}].}

\nomenclature[P]{$t$}{The time horizon is divided into equal-length time periods t, which should be aligned with the electrical markets.}
\nomenclature[P]{$\mathcal T$}{Number of periods on a given time horizon the model is optimizing.}
\nomenclature[P]{$\mathcal N$}{Total number of silos involved in the plant.}
\nomenclature[P]{$\mathcal K$}{Total number of electrical machines involved in the plant.}
\nomenclature[P]{${P_\mathrm{PV}}_t$}{Power generated by the PV system in the period $t$ [{\unit[per-mode = symbol]{\MW}}].}

\printnomenclature

\bibliography{Bibliografia}
\bibliographystyle{apalike}

\end{document}